\documentclass{optica-article}

\journal{opticajournal} 

\articletype{Research Article}

\usepackage{lineno}

\newcommand{\ums}{\ensuremath{\mu\mathrm{s}}}
\newcommand{\mperp}{\ensuremath{\mathrm{m}^{-2}}}

\begin{document}

\title{Synergy between laser linewidth and frequency chirp in mesospheric
magnetometry based on the sodium laser guide star}

\author{Yucheng Yang,\authormark{1,2,*} Kai Guo,\authormark{3} Chunyang
Lei,\authormark{4} and Shuai Wang\authormark{5}}

\address{\authormark{1}State Key Laboratory of Spatial Datum, Beijing 100020,
China\\
\authormark{2}Beijing Key Laboratory of Quantum Sensing and Precision
Measurement, Center for Quantum Information Technology, and Institute of
Quantum Electronics, Peking University, Beijing 100871, China\\
\authormark{3}Institute of Systems Engineering, AMS, Beijing 100141, China\\
\authormark{4}AthenaEyes Co., Ltd., Changsha 410000, China\\
\authormark{5}Institute of Optics and Electronics, CAS, Chengdu 610209, China}

\email{\authormark{*}yangyucheng@pku.edu.cn}

\begin{abstract*}
Mesospheric sodium magnetometry measures the geomagnetic field near 90~km by
reading a magnetic resonance out of the fluorescence of a laser guide star.
Its sensitivity depends on two spectral parameters of the pump laser, the
linewidth and the frequency chirp rate, which earlier work optimized one at
a time. We scan both jointly, using velocity-resolved density-matrix
simulations of Larmor-synchronous pulsed Na~D$_2$ excitation at the 20\%
duty cycle standard for this format. The two parameters share a single duty:
mitigating photon recoil. When the chirp carries that duty, the linewidth
optimum contracts to the power-broadened homogeneous width and the return
flux stays within 1\% of its peak across 2--10~MHz. The flux-optimal chirp
rate is $0.19~\mathrm{MHz}/\ums$, one fifth of the continuous-wave value, and
it lies 36\% above the matched recoil-tracking rate; away from it the
linewidth optimum shifts by $1.8\times$. The spin polarization rises
monotonically over the whole map, so the flux optimum is not the sensitivity
optimum. Contrast, uncomputed here, should displace the latter toward
broader lines.
\end{abstract*}



\section{Introduction}

A layer of atomic sodium vapor, some 10~km thick, sits at an altitude near
92~km. It is the gain medium of modern laser guide star (LGS) adaptive
optics~\cite{Foy,Thompson,Happer}, and more recently it has become the
sensing volume of a remote magnetometer. Circularly polarized 589-nm pump
light prepares the spin of the sodium atoms. That polarization then
precesses about the geomagnetic field. Modulating the pump at the Larmor
frequency imprints a magnetic resonance on the return flux, and the field at
90~km is read off from it~\cite{Higbie,Kane,Bustos2018,Fan}. The scheme is
synchronous optical pumping in the sense of Bell and
Bloom~\cite{Bell,BudkerRomalis}, executed over a 90-km standoff. Geometry
makes it photon-starved: only about $10^{-11}$ of the emitted photons are
collected~\cite{Higbie}. Every design choice that raises the return flux per
atom therefore converts directly into magnetometric sensitivity. That is the
design pressure.

For continuous-wave (CW) astronomical beacons, flux maximization has a
mature literature. It covers irradiance and
polarization~\cite{Holzlohner,Milonni}, pulse
formats~\cite{Bradley,Holzlohner2012,Rampy}, photon
recoil~\cite{Hillman2008}, frequency chirping, realized there as continuous
large-amplitude sweeps of 150--200~MHz across the Doppler
core~\cite{Bustos2020,Hellemeier}, and the laser
linewidth~\cite{Liu2021,LiHY,Bolbasova}. A magnetometer cannot simply import
these answers. Its pump light must be pulsed at the Larmor frequency, and
pulsing changes the optimization qualitatively. This paper fixes the two
spectral parameters of such a pulse train, the laser linewidth $\Delta\nu$
and the chirp rate $\alpha$. We scan them jointly, under the pulse format
the magnetometer is committed to, with the temporal parameters held at the
20\% duty cycle standard for this format~\cite{Higbie}.

The central finding is that the two parameters are coupled. Neither is
optimized in isolation. They are two implementations of a single physical
duty, and the optimum of each depends on how much of that duty the other
already carries. The Doppler width of
mesospheric sodium at 185~K is 621~MHz, while the natural linewidth of the
D$_2$ transition is $\Gamma/2\pi = 9.79$~MHz~\cite{Steck}. A monochromatic
laser therefore addresses roughly one atom in sixty. Broadening the laser
line recruits more velocity classes at once. Chirping the laser frequency
recruits them in sequence, and it also compensates photon recoil, which
shifts an atom's resonance by 50~kHz per scattering
event~\cite{Hillman2008,Bustos2020}. Both knobs have been studied
separately for CW beacons, the linewidth in
Refs.~\cite{Liu2021,LiHY,Bolbasova} and the chirp in
Refs.~\cite{Bustos2020,Hellemeier}. Their joint behavior under
Larmor-synchronous pulsing has not been mapped.

The joint scan gives a precise result. The recoil-mitigation duty is
partitioned between the linewidth and the chirp. Under Larmor-synchronous
pulsing the chirp optimum is set by the recoil drift accumulated over the
bright fraction of the cycle, while the linewidth optimum contracts to the
homogeneous width. The chirp waveform that emerges is not the CW snowplow
at reduced speed. It is a small bounded sawtooth, tens of megahertz in
amplitude, synchronized to the pulse train, with a rate that tracks the
duty-scaled recoil drift.

\begin{figure}[htbp]
\centering
\includegraphics[width=\textwidth]{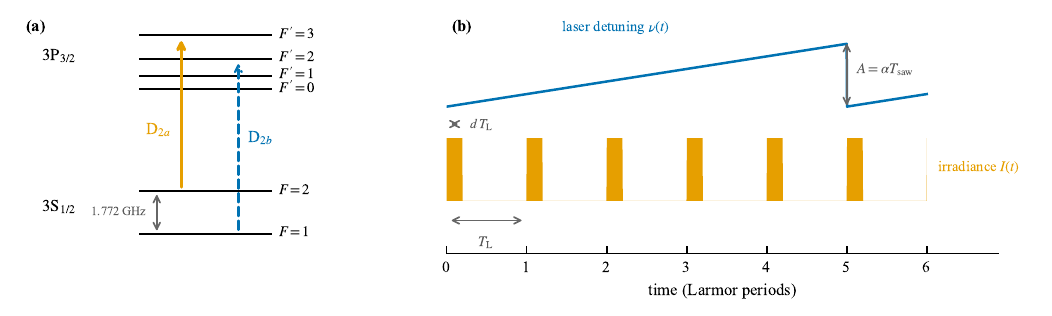}
\caption{The excitation scheme. (a) Hyperfine structure of the Na~D$_2$ line
(spacings not to scale). The circularly polarized pump light drives the
D$_{2a}$ group from $F=2$ toward the cycling transition to $F'=3$; a sideband
on D$_{2b}$ repumps population lost to $F=1$. (b) The pulse train and the
frequency sweep. The irradiance is square-modulated at the Larmor period
$T_{\rm L}$ with duty cycle $d$; the optical frequency rises at the chirp
rate $\alpha$ and resets after $T_{\rm saw}=40$ Larmor periods, giving a
sawtooth of amplitude $A=\alpha T_{\rm saw}$.}
\label{f1}
\end{figure}

\section{Model and simulation conditions}

The simulations solve the optical Bloch equations of the complete Na~D$_2$
hyperfine structure. The model comprises the $3\mathrm{S}_{1/2}$ $F=1,2$
ground manifolds and the $3\mathrm{P}_{3/2}$ $F'=0$--$3$ excited manifolds,
with all 24 Zeeman sublevels. We use the \textit{LGSBloch} extension of the
Atomic Density Matrix framework~\cite{Holzlohner,Holzlohner2012,ADM},
machinery validated against CW and pulsed beacon
photometry~\cite{Holzlohner,Rampy}. The velocity dimension is explicit. The
Doppler distribution is partitioned into 201 spectrum-weighted velocity
groups spanning $\pm 3$ Doppler widths, about 18.6~MHz per group or 1.9
natural linewidths. The convergence study below bounds the error this
discretization leaves. The groups are coupled and relaxed by single-photon
recoil ($\delta\nu_r = 50$~kHz per scattering event), velocity-changing
collisions ($35~\ums$), and spin damping, dominated by spin exchange with
atmospheric O$_2$ ($245~\ums$)~\cite{Holzlohner,Higbie}.

Table~\ref{tab1} collects the simulation conditions. The field parameters
follow from entering the coordinates of the Changping campus of Peking
University ($40.25^{\circ}$N, $116.20^{\circ}$E) into the World Magnetic
Model 2025~\cite{WMM}. The result is $B=0.5506$~G and a Larmor period
$T_{\rm L}=2.595~\ums$. The pump light is circularly polarized on D$_{2a}$
and carries a 12\% repump sideband on D$_{2b}$, near the flux-optimal
fraction of 0.137 found in our preprint study of the repumping
parameters~\cite{Repump}. That preprint addresses the repumping manifold;
the spectral axes scanned here are not treated there. Within the present
framework the sideband has a single role, setting the return rate from
$F=1$, and the maps of Section~3 vary smoothly with it. The irradiance
during the bright window is $50~\mathrm{W/m^2}$, of the order of the
cycling-transition saturation intensity. The duty cycle is held at 20\%, the
value Higbie \textit{et al.}~\cite{Higbie} find optimal for
Larmor-synchronous modulation.

\begin{table}[htbp]
\caption{Simulation conditions. Field values are from WMM2025 at the
Changping campus of Peking University~\cite{WMM}.}
\label{tab1}
\centering
\begin{tabular}{ll}
\hline
Parameter & Value\\
\hline
Temperature & 185~K\\
Geomagnetic field $B$ & 0.5506~G (WMM2025)\\
Field zenith, azimuth & $121.6^{\circ}$, $-7.3^{\circ}$\\
Larmor period $T_{\rm L}$ & $2.595~\ums$\\
Recoil shift per event $\delta\nu_r$ & 50~kHz\\
Velocity-changing collision time & $35~\ums$\\
Spin-damping time & $245~\ums$\\
Irradiance (bright window) & $50~\mathrm{W/m^2}$\\
Duty cycle $d$ & 20\%\\
Repump fraction & 12\%~\cite{Repump}\\
Sawtooth reset period $T_{\rm saw}$ & 40\,$T_{\rm L}$ ($103.8~\ums$)\\
Periods per point & 120 (3 sawtooth cycles)\\
Velocity groups & 201 over $\pm 3$ Doppler widths\\
Linewidth grid $\Delta\nu$ & 0.1--40~MHz, 16 values\\
Chirp-rate grid $\alpha$ & 0.1--1.2~MHz/$\ums$, 12 values\\
\hline
\end{tabular}
\end{table}

The optical frequency follows a bounded sawtooth (Fig.~\ref{f1}). It rises
at the chirp rate $\alpha$ for 40 Larmor periods and then resets, so the
sweep amplitude $A=\alpha\times 103.8~\ums$ stays a small excursion into the
Doppler core. The reset period of $40\,T_{\rm L}$ rests on three grounds. It
keeps the sweep synchronized to an integer number of pulse periods. It caps
the amplitude at 20\% of the Doppler width even at the fastest chirp
scanned, so the sweep never leaves the populous core. At $103.8~\ums$ it
also equals three velocity-changing collision times, so the velocity classes
depleted during one sweep cycle are rethermalized by the time the sweep
returns to them. The laser line is modeled as a comb of spectral components
of total width $\Delta\nu$ (FWHM).

The scanned grid covers $\Delta\nu = 0.1$--$40$~MHz in 16 steps and
$\alpha=0.1$--$1.2~\mathrm{MHz}/\ums$ in 12 equally spaced steps of
$0.1~\mathrm{MHz}/\ums$: 192 velocity-resolved simulations in all. The
$\Delta\nu$ values are not equally spaced. They run 0.1, 0.4, 0.7, 1.0, then
2.0--10.0~MHz in steps of 1~MHz, and finally 20, 30 and 40~MHz, resolving
the region where the optimum lies while bracketing it with decade-scale
points. The flux maximum reported below falls on the 8~MHz grid point.

Each grid point runs 120 Larmor periods, three sawtooth cycles. The reported
quantity is the specific return flux $\Psi$ of beacon
photometry~\cite{Holzlohner,Bustos2020}: the return flux per unit of
period-averaged irradiance, in photons per second, steradian and atom per
$\mathrm{W}\,\mperp$, averaged over the last complete cycle. Irradiance and
duty cycle are fixed throughout this paper, so $\Psi$ is proportional to the
absolute return flux and no comparison below is affected by the
normalization. This fixed-integration-time convention makes all points
strictly comparable. It reads the flux in its cyclo-stationary regime. The
ensemble spin polarization, by contrast, is still evolving at the readout,
so we quote it only as a relative trend (Section~4).

Two systematic checks bound the numerics, and the choice of 201 velocity
groups needs its justification. We repeated the worst-case grid point, the
narrowest linewidth at the fastest chirp, at 101, 201 and 401 groups. At 101
groups the discrete velocity grid imprints a spurious comb on the per-period
flux, with a root-mean-square amplitude of 3.2\% and peaks above 8\%, locked
to the accumulated sweep and repeating in every sawtooth cycle. This
artifact must be excluded, because the grid is rebuilt at every map point
and its residue would vary across the plane. At 201 groups the same residual
falls to 0.22\% root-mean-square, fifteen times smaller, and the windowed
flux carries only a smooth $-0.6\%$ discretization offset that leaves the
map topography unaffected. Refining further to 401 groups shifts the
windowed flux by an additional 0.26\% while multiplying the cost per point
by 5.6, a poor exchange. We therefore use 201 groups and declare the
$-0.6\%$ offset a systematic on absolute $\Psi$.

Reproducibility was checked independently. Two grid points had been computed
three weeks earlier, on a different cluster partition and under an earlier
revision of the driver scripts. Those revisions left the solver, its
tolerances, the atomic data and the initial conditions untouched. The two
runs agree point for point, over all 120 periods, to a relative
$5\times10^{-12}$. That is the level of the solver's own deterministic
reproducibility.

\section{The flux over the two spectral axes}

\begin{figure}[htbp]
\centering
\includegraphics[width=\textwidth]{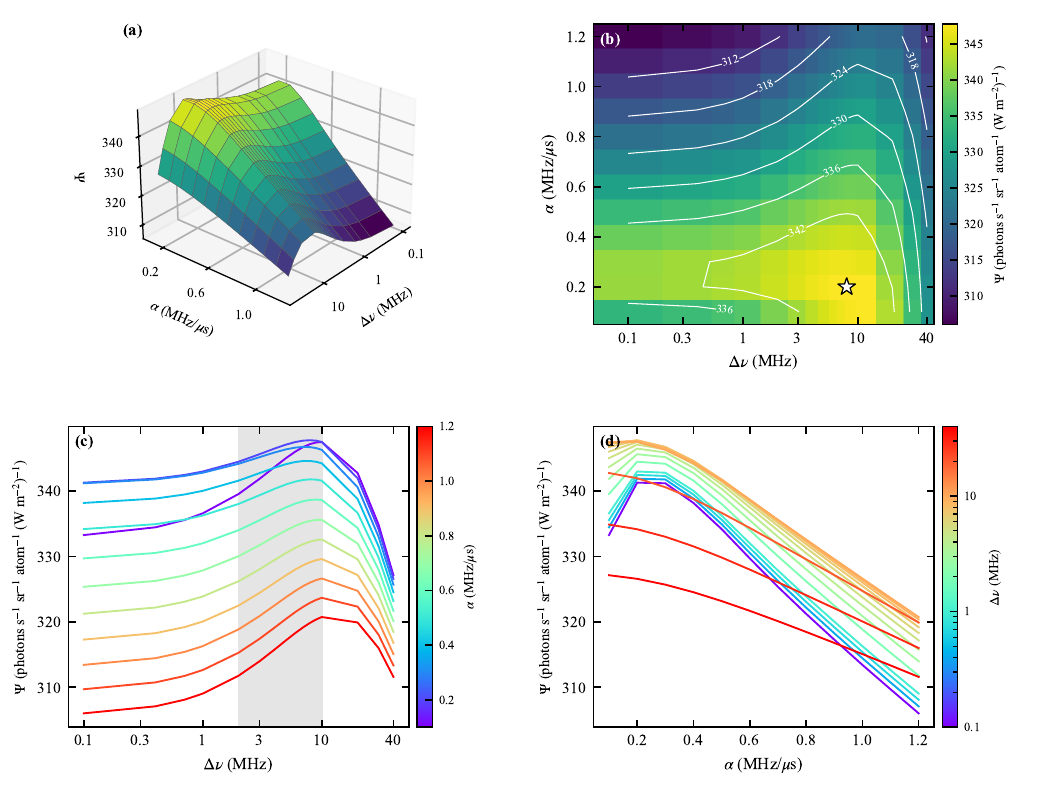}
\caption{Return flux over the complete scanned plane, all 192 grid points.
(a) Surface plot of $\Psi(\Delta\nu,\alpha)$. (b) Top view; contours in the
units of the colorbar, the star marking the maximum $\Psi=347.8$ at
$\Delta\nu=8$~MHz, $\alpha=0.2~\mathrm{MHz}/\ums$. (c) Projection along the
linewidth axis: each curve is one chirp rate, colored by the bar on the
right; the shaded band, 2--10~MHz, holds the flux within 1\% of the maximum
at the optimal chirp. (d) Projection along the chirp axis: each curve is one
linewidth. The flux turnover moves from $0.2~\mathrm{MHz}/\ums$ at narrow
linewidths to the scan edge at the widest, where the broad line already
covers the velocity classes that the sweep would otherwise recruit.}
\label{f2}
\end{figure}

Figure~\ref{f2} presents the full map: the surface, its top view, and both
families of projections, so that every simulated point is visible. The flux
maximum, $\Psi=347.8$ at $(\Delta\nu,\alpha)=(8~\mathrm{MHz},\,
0.2~\mathrm{MHz}/\ums)$, is interior on both axes, and the surrounding
topography is gentle: the worst corner of the entire map lies only 12\%
below the peak. The two axes nevertheless carry different physics, which the
two projection families separate cleanly. We take them in turn.

Consider first the linewidth axis [Fig.~\ref{f2}(c)]. At the flux-optimal
chirp the maximum sits at 8~MHz, but the plateau around it is broad: every
linewidth from 2 to 10~MHz yields a flux within 1\% of the peak. The
linewidth specification is therefore a tolerance band rather than a single
value. The penalties outside the band are modest and asymmetric, $-1.9\%$
at $\Delta\nu=0.1$~MHz and $-6.1\%$ at 40~MHz, and the band shifts only
weakly with the chirp rate: 3--10~MHz at $\alpha=0.6~\mathrm{MHz}/\ums$ and
6--20~MHz at the fastest chirp scanned, where the faster sweep prefers a
broader line to hand atoms from one spectral component to the next.

Both edges of the plateau have clean physical origins. The lower edge is
saturation relief. At $50~\mathrm{W/m^2}$ the pump runs near the saturation
intensity of the cycling transition
($I_{\rm sat}=62.6~\mathrm{W/m^2}$~\cite{Steck}), so a very narrow line
concentrates all its power on velocity classes that are already saturated.
Their fluorescence grows only as $s/(1+s)$ with the saturation parameter
$s$, while neighboring classes remain unaddressed. Broadening the line
toward the homogeneous width recruits those neighbors at near-linear
efficiency, which is why the flux climbs between 0.1 and 2~MHz. The climb
must stop once the line fills the homogeneous linewidth, because the
homogeneous wings of each velocity class already overlap its neighbors. At
our saturation parameter the power-broadened width is
$\Gamma\sqrt{1+s}/2\pi = 13$~MHz, and the measured plateau top, 8--10~MHz,
sits just below it.

The upper edge combines two costs, both of which grow with bandwidth. The
first is spectral dilution: beyond the homogeneous width, additional
bandwidth no longer recruits fresh atoms, it only thins the spectral density
seen by each class. The second is downpumping. The $F'=2$ level lies
58~MHz below the cycling transition~\cite{Steck}, and excitation to it
branches on decay to the dark $F=1$ ground state. For the narrow lines of
the plateau this channel is held to the off-resonant floor, suppressed by
$(\Gamma/2\delta)^2 \approx 0.7\%$ of the resonant rate at the 58-MHz
detuning. As the line broadens toward that splitting, an increasing share of
the spectral weight sits at reduced detuning from $F'=2$. The suppression
weakens, and the 12\% repump sideband can no longer keep pace. Together the
two costs produce the steepening decline visible beyond 20~MHz.

The scan carries an internal signature of this division of labor. The
flux-optimal linewidth is not monotonic in the chirp rate. Parabola
refinement along the linewidth axis places it at 9.8~MHz at
$\alpha=0.1~\mathrm{MHz}/\ums$, at a minimum of 7.2~MHz at $\alpha=0.3$,
and back up to 12.8~MHz at $\alpha=1.2$. The minimum sits where the chirp
performs the recoil refresh most completely, near twice the
matched-tracking rate derived below. It lies above rather than at the
matched rate for the same reason the flux optimum does: a sweep
systematically faster than tracking adds fresh-class coverage at little
tracking cost, so the total refresh delivered by the sweep peaks somewhat
above matching. On either side of the minimum, when the sweep is too slow to
track or so fast that it outruns the atoms, the line is pulled broader to
take back part of the refresh duty. The zero-chirp limit lies outside the
scanned domain, and deliberately so. Even our slowest chirp,
$0.1~\mathrm{MHz}/\ums$, already supplies most of the
$0.14~\mathrm{MHz}/\ums$ recoil drift. No point of this map therefore probes
the regime in which the linewidth must carry the refresh alone.

That regime is documented in the literature, and the comparison places our
band in a consistent physical sequence. For CW beacons without chirping, Liu
\textit{et al.} found that broadening from 0 to anywhere in 1--100~MHz
mitigates recoil and raises the return~\cite{Liu2021}. For QCW pulses
without chirping, Li \textit{et al.} locate the optimum at 60~MHz for a
single-mode line and near 300~MHz for a multimode one~\cite{LiHY}. In both
schemes the linewidth does two jobs at once, covering velocity classes
\textit{and} refreshing recoil-shifted atoms, and the optimum is
correspondingly broad. In the magnetometer scheme of Higbie
\textit{et al.} the optimal linewidth is broader still, 400~MHz on the
D$_1$ line at the same 20\% duty cycle~\cite{Higbie}, because there the
broad line performs a third duty: once most velocity classes are addressed,
a velocity-changing collision no longer removes an atom from resonance, so
broadening directly suppresses the dominant depolarization channel of that
scheme. Our 2--10~MHz band is the opposite endpoint of this sequence. With
the sawtooth chirp performing the recoil refresh, the linewidth retains only
the coverage duty, and its optimum contracts to the homogeneous width. (The
GHz-scale ``linewidths'' of Ref.~\cite{Bolbasova} span the whole D$_2$
manifold, D$_{2b}$ included, and so act as built-in repumping, whereas
$\Delta\nu$ here broadens the D$_{2a}$ pump line alone.) The optimal
linewidth of a sodium-layer instrument is therefore not a universal number
but an accounting of which duties the line must carry. This division of
labor is the synergy of the title, and it argues that a chirped system
should not also carry a wide line, since the wide line's costs are then paid
without its recoil benefit being needed.

Consider next the chirp axis [Fig.~\ref{f2}(d)]. There the optimum is set by
recoil compensation under the constraint of the pulse format. At the
flux-maximal linewidth the flux rises to a maximum at
$\alpha\approx0.2~\mathrm{MHz}/\ums$. Parabola refinement of the three
points around the peak places it at $0.19~\mathrm{MHz}/\ums$, after which
the flux declines by 8\% out to $1.2~\mathrm{MHz}/\ums$. The optimum is the
rate at which the sweep keeps pace with the recoiling atoms. An atom
scattering resonant photons recoils by 50~kHz per event, and at our
irradiance and duty cycle its period-averaged resonance drift is
$d\,(\Gamma/2)\,[s/(1+s)]\,\delta\nu_r = 0.14~\mathrm{MHz}/\ums$, where
$\Gamma/2 = 30.8\times10^{6}~\mathrm{s^{-1}}$ is half the radiative decay
rate of the excited state ($\Gamma/2\pi = 9.79$~MHz). A chirp near this rate
drags the addressed atoms along in frequency space. That is the snowplow
effect of CW chirped-beacon theory~\cite{Bustos2020,Hellemeier}, throttled
here by the fact that under Larmor-synchronous pulsing the atoms scatter
only a fraction $d$ of the time.

The measured optimum lies 36\% above the zeroth-order matched rate and 22\%
above the duty-scaled continuous-wave value. The sign is the expected one: a
sweep faster than tracking buys fresh velocity classes while still carrying
the atoms it has. The size is not something the tracking argument supplies.
That argument fixes the rate at which the sweep keeps pace with the drift;
it says nothing about how much fresh-atom coverage is worth buying. The
premium is therefore an empirical feature of the map. The quantitative
anchor is the matched rate itself. For CW excitation at the saturation
intensity, Ref.~\cite{Bustos2020} derives a recoil-accumulation rate of
$0.78~\mathrm{MHz}/\ums$ and finds the CW optimum at
0.8--$1.0~\mathrm{MHz}/\ums$. Scaling that matched rate by our 20\% duty
cycle gives $0.156~\mathrm{MHz}/\ums$, between our predicted 0.14 and
measured 0.19. Only the \textit{rate} inherits this duty-cycle scaling. The
sweep \textit{amplitude}, $A=\alpha T_{\rm saw}\approx20$~MHz here, is fixed
by the designed reset period rather than by the duty cycle, and is an order
of magnitude below the 150--200-MHz sweeps of the CW
experiments~\cite{Bustos2020,Hellemeier}: the pulsed optimum is a small
synchronized sawtooth, not the CW snowplow slowed down.

The result also carries a site dependence worth stating. Because the reset
period is fixed at 40 Larmor periods, $T_{\rm saw}=40\,T_{\rm L}\propto1/B$.
At a given chirp rate the sweep amplitude $A=\alpha T_{\rm saw}$ therefore
scales as $\alpha/B$ and moves with the observing site. The recoil-tracking
rate itself is set by atomic scattering and is independent of $B$ to zeroth
order. The scaling is demonstrated at $50~\mathrm{W/m^2}$; since the CW
matched rate itself grows with irradiance, from 0.7 to
$2.0~\mathrm{MHz}/\ums$ in Ref.~\cite{Bustos2020}, higher-power operation
requires re-evaluating the optimum. At the widest linewidths the chirp
optimum slides to the low edge of the scanned range, visible in the lightest
curves of Fig.~\ref{f2}(d). A 40-MHz line already spans many velocity
classes, so sweeping adds coverage the line owns anyway while still paying
the recoil-mismatch cost. Broad lines want slow chirps, one more expression
of the two knobs sharing a single job.

The practical summary of both axes is a specification with wide margins: any
linewidth from 2 to 10~MHz, and a chirp rate near $0.2~\mathrm{MHz}/\ums$ at
the 20\% duty cycle, with single-percent penalties for missing either target
by a factor of two.

\section{Spin polarization and magnetometric sensitivity}

\begin{figure}[htbp]
\centering
\includegraphics[width=\textwidth]{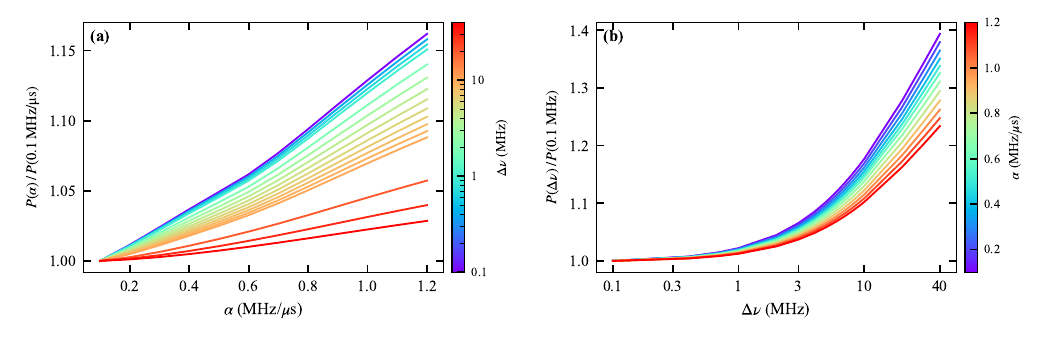}
\caption{Ensemble spin polarization across the map, as relative trends.
(a) Against the chirp rate, normalized to
$\alpha=0.1~\mathrm{MHz}/\ums$, one curve per linewidth. (b) Against the
linewidth, normalized to $\Delta\nu=0.1$~MHz, one curve per chirp rate. The
rise is monotonic along both axes at every value of the other parameter;
there is no interior optimum anywhere on the map. Values are fixed-window
readouts of a still-equilibrating quantity, so only the trends are
meaningful.}
\label{f3}
\end{figure}

A magnetometer values spin polarization alongside photon number, for a
quantitative reason, and that reason fixes how this section should be read.
For an ensemble of $N$ uncorrelated atoms with gyromagnetic ratio $\gamma$,
the spin-projection-noise limit on a measurement of duration $T$ with
coherence time $\tau$ is
$\delta B\approx1/(\gamma\sqrt{NT\tau})$~\cite{BudkerRomalis}. A remote
sodium magnetometer operates far from this limit. With a collection fraction
of order $10^{-11}$~\cite{Higbie}, the readout is photon-shot-noise limited
and the sensitivity scales as $\delta B\propto1/(\gamma\,C\sqrt{RT})$, where
$R$ is the detected photon rate and $C$ the contrast of the Larmor resonance
in the modulated return~\cite{Kane,NSR}. The two factors divide cleanly
between what this campaign determines and what it does not. $R$ is
determined exactly: it is proportional to the return flux $\Psi$ of
Section~3, times receiver constants. $C$ is not. Evaluating it requires the
resonance lineshape, obtained by sweeping the modulation frequency across
the Larmor frequency, whereas every simulation here drives the atoms exactly
on resonance. That is the gap. What the present maps supply for $C$ is its
atomic ingredient,
the ensemble spin polarization, which sets how deeply the Larmor precession
can modulate the fluorescence. We treat $C$ as monotone in that polarization
at fixed receiver geometry, and we flag this as the one link in the
sensitivity chain that the present data set asserts rather than computes.
Under it, the spectral specification feeds the sensitivity through
$\sqrt{\Psi}$ and the polarization through the contrast, the two observables
entering as the product $C\sqrt{R}$.

The two factors enter with different powers, and that asymmetry already
fixes the direction in which the sensitivity optimum moves. Across the
2--10~MHz band the flux varies by less than 1\%, so $\sqrt{R}$ varies by
less than 0.5\%. Over the same band, read with the same fixed window, the
spin polarization rises by 12\% at the flux-optimal chirp
[Fig.~\ref{f3}(b), the $\alpha=0.2~\mathrm{MHz}/\ums$ curve over
2--10~MHz]. Because $C$ enters $\delta B$ to the first power and $R$ only to
the half power, the contrast term dominates the photon term, and $\delta B$
falls monotonically toward the broad edge of the band: under the
monotonicity assumption the sensitivity optimum is not at the flux optimum
but lies at or beyond 10~MHz. The chirp axis reads the same way. Between the
flux optimum at $0.19~\mathrm{MHz}/\ums$ and the edge of the scan at
$1.2~\mathrm{MHz}/\ums$ the flux loses 8\%, hence $\sqrt{R}$ about 4\%,
while the polarization at the flux-optimal linewidth gains 8.7\%
[Fig.~\ref{f3}(a), the $\Delta\nu=8$~MHz curve between $\alpha=0.2$ and
$1.2~\mathrm{MHz}/\ums$], so $\delta B$ falls by about 4\% as well. The
trend cannot run indefinitely. The polarization rises only while fresh
velocity classes remain to be recruited, so $C$ must eventually fall, and
the 400~MHz optimum that Higbie \textit{et al.} obtained by optimizing the
sensitivity directly is consistent with this~\cite{Higbie}. The upper edge
of the flux plateau is set in any case by overlap with the $F'=2$
downpumping resonance, which depresses the achievable spin orientation. We
therefore expect the sensitivity optimum near or just beyond the broad end
of the plateau. This conclusion rests on the uncomputed contrast and is
offered as a bound on the direction of the displacement, not as a prediction
of its magnitude. What the present maps do establish is that a magnetometer
specified on the flux maximum alone would be biased toward too narrow a line
and too slow a chirp.

Figure~\ref{f3} shows how the spin polarization responds to the two spectral
parameters. It has not reached steady state within the integration window.
Its equilibration is paced by the 245-$\ums$ spin-damping time, 2.4 sweep
cycles long, while the window covers $120\,T_{\rm L}=311~\ums$, about
$1.27\,\tau_d$; this interval is identical at every grid point, so the
comparison across the map is unaffected. We therefore quote only its trend,
which is unambiguous: it rises monotonically along both axes, by 16\% across
the chirp range at the narrowest line [Fig.~\ref{f3}(a)] and by up to 39\%
across the linewidth range at the slowest chirp [Fig.~\ref{f3}(b)]. These
percentages are fixed-window readouts whose amplitudes will grow toward
equilibrium, while the monotonic ordering is the calibrated statement. There
is no interior maximum anywhere on the map.

Carrying this ranking over to equilibrium is a short step. All points start
from the same unpolarized ensemble and relax with the same 245-$\ums$
spin-damping time. To the accuracy of a single-scalar description they
differ only through an effective pumping rate $\Lambda$. Since both the
steady value and the rate at which it is approached grow with $\Lambda$, the
polarization read at any fixed elapsed time ranks the points as their steady
states do. The reduction itself remains an assumption---a point that opens
the $F'=2$ downpumping channel is not exactly a point with a larger
$\Lambda$---and the residual is not quantified here. The reading is
straightforward. Spin polarization is a per-atom inventory, and even the
largest sweep amplitude in the map ($A=125$~MHz at
$\alpha=1.2~\mathrm{MHz}/\ums$) together with the widest linewidth (40~MHz)
addresses well under a fifth of the 621-MHz Doppler distribution. Every
increment of sweep or linewidth therefore still recruits new atoms into the
polarized population. On these axes the scanned box ends before the
trade-off does.

Two consequences follow for the magnetometer. First, the flux and the spin
polarization prefer different spectral settings: the flux turns over at
$\alpha=0.2~\mathrm{MHz}/\ums$ and $\Delta\nu=8$~MHz, while the
polarization is still rising at $1.2~\mathrm{MHz}/\ums$ and 40~MHz. Through
the sensitivity product $C\sqrt{R}$, the operating point that minimizes
$\delta B$ therefore lies between the flux optimum and the
polarization-preferred corner of the map---toward the broad end, as the
power counting above indicates---and locating it exactly requires the
contrast calibration of a specific receiver rather than the atomic response
alone. These are predictions of a simulation and await experimental test. In
operation they are testable in one campaign, by scanning $\alpha$ and
$\Delta\nu$ while recording both the mean return and the Larmor-resonance
contrast, which would verify the flux map and calibrate $C$ simultaneously.
Such a campaign requires a 589-nm source with adjustable linewidth and a
programmable frequency sweep, and it must resolve relative changes in the
return-photon rate at the 1\% level, the scale of the plateau. The same
separation appears on the repump-fraction axis, where the two optima differ
by a factor of 2.5--2.8~\cite{Repump}: in this excitation scheme brightness
and spin polarization are maximized by different drives. Second, because on
this plane the sweep amplitude is tied to the chirp rate
($A=\alpha T_{\rm saw}$ at fixed $T_{\rm saw}$), the monotonic polarization
rise does not by itself distinguish a faster sweep from a wider one. The
flux optimum is immune to this ambiguity, recoil tracking being a statement
about rate rather than width; the polarization trend should be read as a
coverage effect pending a fixed-amplitude scan.

\section{Conclusions}

Velocity-resolved density-matrix simulations of the Na~D$_2$ line under
Larmor-synchronous pulsed excitation give the spectral half of the laser
specification for a mesospheric magnetometer based on the sodium guide star.
Its central property is a wide margin. At the flux-optimal chirp the laser
linewidth is a 2--10~MHz tolerance band, bounded below by saturation relief
that completes at the power-broadened homogeneous width and above by
spectral overlap with the $F'=2$ downpumping resonance, inside which the
return flux varies by less than 1\%. The chirp rate has a genuine interior
optimum at the duty-cycle-scaled recoil-tracking rate,
$0.19~\mathrm{MHz}/\ums$ at the 20\% duty cycle, consistent with the
duty-scaled CW recoil-accumulation rate of Ref.~\cite{Bustos2020} to within
the premium just quantified; missing it by a factor of two costs single
percent.

Comparison with the CW linewidth literature~\cite{Liu2021} resolves an
apparent discrepancy and names the synergy: linewidth broadening and
frequency chirping are two implementations of the same recoil-mitigation
duty, and when the chirp carries that duty the linewidth optimum contracts
to the homogeneous width. The ensemble spin polarization rises
monotonically over the entire map, a coverage effect, so brightness and spin
polarization favor different operating points. Through the
photon-shot-noise sensitivity $\delta B\propto1/(C\sqrt{R})$ this
difference is not a tie, because the contrast enters to the first power and
the photon rate only to the half power. Within the 2--10~MHz band the
polarization gains 12\% while $\sqrt{R}$ moves by less than 0.5\%, so the
two optima cannot coincide and the sensitivity optimum is displaced at least
to the broad end of the plateau. Locating it exactly requires the contrast
calibration of a specific receiver rather than the atomic response alone.

\begin{backmatter}
\bmsection{Funding}
National Natural Science Foundation of China (Grant No.~62301377).

\bmsection{Disclosures}
The authors declare no conflicts of interest.

\bmsection{Data availability}
Data underlying the results presented in this paper are not publicly
available at this time but may be obtained from the authors upon reasonable
request.
\end{backmatter}

\bibliography{refs}

@article{Foy,
  author = {R. Foy and A. Labeyrie},
  title = {Feasibility of adaptive telescope with laser probe},
  journal = {Astron. Astrophys.},
  volume = {152},
  pages = {L29},
  year = {1985},
  url = {https://articles.adsabs.harvard.edu/pdf/1985A%26A...152L..29F}
}

@article{Thompson,
  author = {L. A. Thompson and C. S. Gardner},
  title = {Experiments on laser guide stars at Mauna Kea Observatory for adaptive imaging in astronomy},
  journal = {Nature},
  volume = {328},
  pages = {229},
  year = {1987},
  doi = {10.1038/328229a0}
}

@article{Happer,
  author = {W. Happer and G. J. MacDonald and C. E. Max and F. J. Dyson},
  title = {Atmospheric turbulence compensation by resonant optical backscattering from the sodium layer in the upper atmosphere},
  journal = {J. Opt. Soc. Am. A},
  volume = {11},
  pages = {263},
  year = {1994},
  doi = {10.1364/JOSAA.11.000263}
}

@article{Higbie,
  author = {J. M. Higbie and S. M. Rochester and B. Patton and R. Holzl{\"o}hner and D. Bonaccini Calia and D. Budker},
  title = {Magnetometry with mesospheric sodium},
  journal = {Proc. Natl. Acad. Sci. USA},
  volume = {108},
  pages = {3522},
  year = {2011},
  doi = {10.1073/pnas.1013641108}
}

@article{Kane,
  author = {T. J. Kane and P. D. Hillman and C. A. Denman and M. Hart and R. P. Scott and M. E. Purucker and S. J. Potashnik},
  title = {Laser remote magnetometry using mesospheric sodium},
  journal = {J. Geophys. Res. Space Phys.},
  volume = {123},
  pages = {6171},
  year = {2018},
  doi = {10.1029/2018JA025178}
}

@article{Bustos2018,
  author = {F. Pedreros Bustos and D. Bonaccini Calia and D. Budker and M. Centrone and J. Hellemeier and P. Hickson and R. Holzl{\"o}hner and S. Rochester},
  title = {Remote sensing of geomagnetic fields and atomic collisions in the mesosphere},
  journal = {Nat. Commun.},
  volume = {9},
  pages = {3981},
  year = {2018},
  doi = {10.1038/s41467-018-06396-7}
}

@article{Fan,
  author = {T. Fan and X. Yang and J. Dong and L. Zhang and S. Cui and J. Qian and R. Dong and K. Deng and T. Zhou and K. Wei and Y. Feng and W. Chen},
  title = {Remote magnetometry with mesospheric sodium based on gated photon counting},
  journal = {J. Geophys. Res. Space Phys.},
  volume = {124},
  pages = {7505},
  year = {2019},
  doi = {10.1029/2019JA026956}
}

@article{Bell,
  author = {W. E. Bell and A. L. Bloom},
  title = {Optically driven spin precession},
  journal = {Phys. Rev. Lett.},
  volume = {6},
  pages = {280},
  year = {1961},
  doi = {10.1103/PhysRevLett.6.280}
}

@article{BudkerRomalis,
  author = {D. Budker and M. Romalis},
  title = {Optical magnetometry},
  journal = {Nat. Phys.},
  volume = {3},
  pages = {227},
  year = {2007},
  doi = {10.1038/nphys566}
}

@article{Holzlohner,
  author = {R. Holzl{\"o}hner and S. M. Rochester and D. Bonaccini Calia and D. Budker and J. M. Higbie and W. Hackenberg},
  title = {Optimization of cw sodium laser guide star efficiency},
  journal = {Astron. Astrophys.},
  volume = {510},
  pages = {A20},
  year = {2010},
  doi = {10.1051/0004-6361/200913108}
}

@article{Milonni,
  author = {P. W. Milonni and H. Fearn and J. M. Telle and R. Q. Fugate},
  title = {Theory of continuous-wave excitation of the sodium beacon},
  journal = {J. Opt. Soc. Am. A},
  volume = {16},
  pages = {2555},
  year = {1999},
  doi = {10.1364/JOSAA.16.002555}
}

@article{Bradley,
  author = {L. C. Bradley},
  title = {Pulse-train excitation of sodium for use as a synthetic beacon},
  journal = {J. Opt. Soc. Am. B},
  volume = {9},
  pages = {1931},
  year = {1992},
  doi = {10.1364/JOSAB.9.001931}
}

@article{Holzlohner2012,
  author = {R. Holzl{\"o}hner and S. M. Rochester and D. Bonaccini Calia and D. Budker and T. Pfrommer and J. M. Higbie},
  title = {Simulations of pulsed sodium laser guide stars: an overview},
  journal = {Proc. SPIE},
  volume = {8447},
  pages = {84470H},
  year = {2012},
  doi = {10.1117/12.924882}
}

@article{Rampy,
  author = {R. Rampy and D. Gavel and S. M. Rochester and R. Holzl{\"o}hner},
  title = {Toward optimization of pulsed sodium laser guide stars},
  journal = {J. Opt. Soc. Am. B},
  volume = {32},
  pages = {2425},
  year = {2015},
  doi = {10.1364/JOSAB.32.002425}
}

@article{Hillman2008,
  author = {P. D. Hillman and J. D. Drummond and C. A. Denman and R. Q. Fugate},
  title = {Simple model, including recoil, for the brightness of sodium guide stars created from CW single frequency fasors and comparison to measurements},
  journal = {Proc. SPIE},
  volume = {7015},
  pages = {70150L},
  year = {2008},
  doi = {10.1117/12.790650}
}

@article{Bustos2020,
  author = {F. Pedreros Bustos and R. Holzl{\"o}hner and S. Rochester and D. Bonaccini Calia and J. Hellemeier and D. Budker},
  title = {Frequency chirped continuous-wave sodium laser guide stars: modeling and optimization},
  journal = {J. Opt. Soc. Am. B},
  volume = {37},
  pages = {1208},
  year = {2020},
  doi = {10.1364/JOSAB.389007}
}

@article{Hellemeier,
  author = {J. Hellemeier and M. Enderlein and M. Hager and D. Bonaccini Calia and R. L. Johnson and F. Lison and M. O. Byrd and L. A. Kann and M. Centrone and P. Hickson},
  title = {Laser guide star return-flux gain from frequency chirping},
  journal = {Mon. Not. R. Astron. Soc.},
  volume = {511},
  pages = {4660},
  year = {2022},
  doi = {10.1093/mnras/stac343}
}

@article{Liu2021,
  author = {X. Liu and X. Qian and R. He and D. Liu and C. Cui and C. Fan and H. Yuan},
  title = {Effects of linewidth broadening method on recoil of sodium laser guide star},
  journal = {Atmosphere},
  volume = {12},
  pages = {1315},
  year = {2021},
  doi = {10.3390/atmos12101315}
}

@article{LiHY,
  author = {H.-Y. Li and L. Feng and Q. Bian and M. Li and B.-T. Sun and C. Wang and M. Wang and Y. Liang and R.-T. Wang and J.-W. Zuo and Y. Bo and K. Wei and Z.-X. Shen and Y.-P. Li and S.-J. Xue},
  title = {Numerical study on the influence of the linewidth of a QCW pulsed sodium laser on the brightness of a guide star},
  journal = {Opt. Express},
  volume = {29},
  pages = {40397},
  year = {2021},
  doi = {10.1364/OE.443293}
}

@article{Bolbasova,
  author = {L. A. Bolbasova and S. A. Ermakov and V. P. Lukin},
  title = {Laser linewidth effect on sodium laser guide star brightness in midlatitude atmosphere},
  journal = {Atmos. Ocean. Opt.},
  volume = {37},
  pages = {919},
  year = {2024},
  doi = {10.1134/S1024856024701185}
}

@misc{Steck,
  author = {D. A. Steck},
  title = {Sodium D line data},
  howpublished = {available at http://steck.us/alkalidata},
  year = {2026},
  note = {(August 2026)}
}

@misc{ADM,
  author = {S. M. Rochester},
  title = {AtomicDensityMatrix package},
  howpublished = {available at http://rochesterscientific.com/ADM/},
  year = {2026},
  note = {(August 2026)}
}

@misc{WMM,
  author = {{NOAA National Centers for Environmental Information} and {British Geological Survey}},
  title = {World Magnetic Model 2025},
  howpublished = {calculator available at http://www.geomag.bgs.ac.uk/data\_service/models\_compass/wmm\_calc.html},
  year = {2026},
  note = {(August 2026)}
}

@misc{Repump,
  author = {Y. Yang and C. Lei and K. Guo and S. Wang},
  title = {Optimization of the repumping parameters for a sodium laser guide star magnetometer},
  howpublished = {available at https://doi.org/10.48550/arXiv.2608.08983},
  year = {2026},
  note = {(August 2026)}
}

@article{NSR,
  author = {L. Lei and T. Wu and H. Guo},
  title = {Sensitivity of quantum magnetic sensing},
  journal = {Natl. Sci. Rev.},
  volume = {12},
  pages = {nwaf129},
  year = {2025},
  doi = {10.1093/nsr/nwaf129}
}

\end{document}